\documentclass[superscriptaddress,longbibliography,amsmath,amssymb,nofootinbib]{revtex4-2}
\usepackage{tikz}
\usepackage{pgfplots}
\usepackage{graphicx}
\usepackage{dcolumn}
\usepackage{bm}
\usepackage{esint}
\usepackage{color}
\usepackage{dsfont}

\def\be{\begin{equation}}
\def\ee{\end{equation}}
\usepackage{hyperref}

\usepackage{caption}
\usepackage{subcaption}

\newcommand{\ket}[1]{\vert{#1}\rangle}
\newcommand{\bra}[1]{\langle{#1}\vert}
\newcommand{\braket}[2]{\langle #1 \vert #2 \rangle}
\newcommand{\avg}[1]{\langle{#1}\rangle}

\newcommand{\tr}[1]{\mathrm{tr}\left({#1}\right)}
\newcommand{\del}{\partial}
\newcommand{\intinf}{\int\limits_{-\infty}^{+\infty}}

\newcommand{\Ham}{\mathcal{H}}
\newcommand{\sgn}{{\rm sgn}}

\newcommand{\bfl}{\boldsymbol{\lambda}}
\newcommand{\bfm}{\boldsymbol{\mu}}
\newcommand{\bfbm}{\boldsymbol{\bar{\mu}}}

\DeclareMathOperator*\Dprod{%
\mathchoice
  {\ooalign{$\bullet$\cr\hidewidth$\displaystyle\prod$\hidewidth\cr}}%
  {\ooalign{\scalebox{.7}{$\bullet$}\cr\hidewidth$\textstyle\prod$\hidewidth\cr}\mkern6mu}%
  {\mkern6mu\ooalign{\scalebox{.6}{$\bullet$}\cr\hidewidth$\scriptstyle\prod$\hidewidth\cr}\mkern3mu}%
  {\mkern4mu\ooalign{\scalebox{.5}{$\bullet$}\cr\hidewidth$\scriptscriptstyle\prod$\hidewidth\cr}\mkern2mu}}

\begin{document}

\title{Two-spinon effects on the thermal Tonks-Girardeau gas
}
\author{Felipe Taha Sant'Ana}
\email[Correspondence to: ]{ftaha@ifpan.edu.pl}
\affiliation{
Institute of Physics, Polish Academy of Sciences, Aleja Lotników 32/46, 02-668 Warsaw, Poland
}%

\author{Hui Liu}
\affiliation{
Institute of Physics, Polish Academy of Sciences, Aleja Lotników 32/46, 02-668 Warsaw, Poland
}%


\begin{abstract}
We study the effects of the two-spinon excitations on the field-field correlator of the Tonks-Girardeau gas.   While these excitations have been previously examined in the ground state of the system, their role at finite temperatures remains unexplored. Here, we extend the analysis to the one-dimensional interacting Bose gas at thermal equilibrium, focusing on the one-body correlation function of the infinitely repulsive Lieb-Liniger model. We demonstrate that two-spinon excitations, characterized by two holes within the rapidity distribution, constitute the dominant contribution to the field-field correlator at low temperatures. Furthermore, we analytically show that incorporating additional particle-hole excitations diminishes their contribution, highlighting the efficacy of the two-spinon framework in capturing the essential physics of the system. Numerical evaluations of both the Fredholm determinant and the spectral sum stemming from the two-spinon program, with the addition of particle-hole excitations, reveal convergence at low temperatures. 

\end{abstract}

\maketitle

\section{Introduction}

Interacting quantum systems restricted to move in one spatial dimension compose a significant area of research both in theoretical and experimental physics due to the rich underlying nature of local interparticle interactions and to a more realistic description of physical systems where the fundamentals constituents are fated to interact with each other. These models have found applications in condensed matter physics, statistical mechanics, and quantum field theory, particularly in the study of one-dimensional systems. One of the key goals in such models is to compute physical observables like correlation functions and dynamical response functions, which encode the behaviour of the system at different energy scales and provide a bridge between theory and experiment. For the experimental part of its importance, we have been seeing an increasing interest by cold atoms experiments \cite{PhysRevLett.91.250402,Kinoshita2004,Paredes2004,PhysRevA.85.023623,Kinoshita2006,Meinert2017}. Regarding the theoretical attention and advances on the field, there is an extensively bibliography one can find, but to cite a few we may refer to different techniques employed over the year that consist a fundamental basis for its research, such as perturbation theories \cite{Massignan2014,Burovski2014,PhysRevA.89.063627,PhysRevE.90.032132,PhysRevB.101.104503} whenever the interaction is weak enough to be treated as a perturbative scheme over the noninteracting system, mean-field considerations \cite{Panochko2019,panochko2021static,hryhorchak2021polaron,Koutentakis2021,PhysRevLett.122.183001}, as well as various numerical methods \cite{PROKOF_EV_1993,PhysRevLett.110.015302,PhysRevB.77.020408,grusdt2015new,PhysRevA.95.023619,Grusdt2017}.  

An important feature related to the class of exactly solvable quantum systems known as Bethe ansatz integrable models is the concept of form factors: matrix elements that form the fundamental concept of quantum correlations between particles in a given theory. The computation of form factors in Bethe Ansatz solvable models is notoriously challenging due to the complex structure of their eigenstates. These eigenstates are described by sets of quantum numbers (rapidities) that satisfy the associated Bethe equations. In the thermodynamic limit, the distribution of these rapidities becomes continuous, and form factors must be computed in terms of these continuous distributions. 
Thermodynamic form factors extend the concept of finite-size form factors to the infinite system, where the rapidities are densely packed along continuous distributions. The analytic structure of form factors plays a key role in deriving long-distance asymptotics of correlation functions, which are essential for capturing the physical properties of integrable systems at large scales \cite{Gamayun_2021,Gamayun_2022}. In addition, these form factors provide insight into the dynamics of excitations, the nature of quasiparticles, and the transport properties in such integrable systems. In summary, thermodynamic form factors in Bethe Ansatz integrable models are a powerful tool for understanding the large-scale, low-energy behaviour of quantum systems. They provide exact information about the matrix elements of observables between different eigenstates and are instrumental in deriving correlation functions and studying the physical properties of such models \cite{2018JSMTE..03.3102D,Bootstrap_JHEP,Cortes_Cubero_2020} . 

In this work we focus on the celebrated Lieb-Liniger model \cite{1963_Lieb_PR_130_1,1963_Lieb_PR_130_2} with repulsive interparticle interaction. It describes a one-dimensional system of contact-interacting bosons and has become a staple of low-dimensional and fully integrable physics, which for decades ever continues to find new fields of relevance both theoretically \cite{2016JSMTE..06.4003C,Cazalilla_2011,2011RvMP...83..863P} and experimentally \cite{Kinoshita:2006aa,Hofferberth:2007aa,Cheneau:2012aa,doi:10.1146/annurev-conmatphys-031214-014548,Hung11,Rancon13}. It continues to be a source of insight, perhaps due to the way its underlying simplicity still leads to behaviour that is far from trivial. Despite being one of the exactly solvable models of quantum mechanics, exact results for many observables of interest -- such as spatial and temporal correlations -- are not trivial to extract. Numerical results \cite{PhysRevA.91.043617,PhysRevLett.115.085301,PhysRevA.89.033605} and algorithms such as ABACUS \cite{2009_Caux_JMP_50} are available for finite size cases, as well as alternative approaches using Monte Carlo \cite{Astrakharchik_1,Astrakharchik_2} or phase-space methods \cite{PhysRevA.79.043619}. Still, a detailed understanding of the few-body processes responsible has only recently started to be explicitly formulated for the simplest zero temperature case in terms of intermediary states \cite{Panfil_2021}. 
Determining such ingredients of the full quantum solution has much promise for both a better understanding of the physics of the system and for adaptation to weakly non-integrable models close to the original model, such as gases in a non-uniform energy landscape, or under time-dependent perturbation \cite{1998_Olshanii_PRL_81,Cheneau:2012aa,PhysRevA.89.033601,Caux06,Caux07,PhysRevA.99.023620,Zill11,Zill_2016,Hung11,Rancon13}. Its thermodynamic extension from the ground state \cite{1969_Yang_JMP_10} provides a theoretical framework for solving the model at finite temperatures, which is our main goal here. Besides, it is well known that the Bethe ansatz equations for the Lieb-Liniger model at finite coupling needs numerical approaches to be solvable. A simplification of the model happens at infinite repulsive interactions, also known as Tonks-Girardeau gas \cite{1936_Tonks,1960_Girardeau}. In such a regime, the particles become hard-core bosons, behaving as fermions due the the strong repulsive interaction, which acts analogously as the Pauli exclusion principle, a phenomenon known as \emph{fermionization} \cite{santana_2020}.  It make possible to map the many-body wavefunction of the bosonic system into a fermionic-like one, i.e. in terms of Slater determinant \cite{felipe_2019}, allowing for the calculation of physical quantities of interest. Moreover, the Bethe equations decouple into a simpler and analytically solvable format. The Tonks-Girardeau gas has been experimentally realized in ultra-cold atomic systems, where advances in trapping and manipulating atoms in optical lattices and waveguides allow for the creation of effectively one-dimensional systems. By tuning the interaction strength between atoms using Feshbach resonances or adjusting the dimensionality of the trap, researchers can reach the Tonks-Girardeau limit. These experiments have confirmed the theoretical predictions, showing behaviour like fermionization and suppressed collisions. Due to this the aforementioned analytical simplification together with its experimental importance, we focus this work over such a regime. 

Canonically, elementary excitations of the Lieb-Liniger model can be classified as particle (Lieb type I) and hole (Lieb type II) modes, which are created by removing a particle from the Fermi sea or adding a particle over the Fermi sea, respectively. An arbitrary excited state of the Lieb Liniger model can be described by appropriate stacking of these excitations. These types of excitations form the basis of the Bethe ansatz formalism for finding the exact many-body solution of the uniform system. Regarding the evaluation of expectation value of local operators, or more specifically the average of different combinations of the bosonic field operators, i.e. correlator, can be expressed in terms of a spectral sum over the whole Hilbert space. If we were able to perform the full spectral sums, the correlations would be solved. Unfortunately, the sum becomes intractably large since the Hilbert space grows exponentially with the number of particles.  Therefore, by focusing on a wisely chosen subset of excitations relevant for our observable, we can hope to evaluate it sufficiently well without performing the whole sum. This is why identification of relevant excitations is central to the efficacy of the thermodynamic form factor approach. Once one understands what are the relevant excitations one can restrict the spectral sum to those. A recent discovery is that specific combinations of these elementary excitations accounts for the majority of correlation phenomena in the system at zero temperature \cite{Panfil_2021}. In particular, excitations composed of two holes within the distribution of rapidities that solve the respective Bethe equations, known also a two-spinon (2sp) excitations, were found to provide the largest contribution to the field-field  correlator of the Lieb-Liniger model. Here, we extend this formalism to finite temperatures. 

Differently from the zero-temperature case, where the ground state is described by the Fermi sea, with the particles occupying symmetric and evenly distributed rapidities on momentum space, at finite temperatures the system does not present such a feature. The addition of temperature breaks the organisational structure of rapidities, compelling us to inspect novel manners to probe the system. Thus, we find that relying on an unequivocal representative thermal state described by a specific rapidities discretisation provided by a counting function supply us with the necessary elements. Hence, we implement an analogous procedure from the zero-temperature regime, namely we employ the two-spinon excitations over the $(N+1)-$particle thermal state represented by our specific rapidities distribution. We then organise the spectral sum regarding this formalism and proceed to demonstrate how it captures the essence of the field-field correlator of the Tonks-Girardeau gas at finite temperatures. 

The manuscript is organised as follows. In Sec.~\ref{sec.theory} we provide the basic ingredients to describe the Lieb-Liniger model in terms of the bosonic field operators as well as its interparticle interaction nature, followed by the thermal equilibrium description and its components. In Sec.~\ref{sec.correlations} we define our main interest, namely the field-field correlator of Lieb-Liniger model at finite temperatures, and the mathematical framework used to study its form factors, followed by its strongly repulsive counterpart. In Sec. \ref{sec.2sp} we introduce the concept of two-spinons and its framework in order to work with the correlator. We show that whenever adding further particle-hole excitations, we end up with a smaller form factor than the one we started with. We then formulate a practical approach to discretise the rapidities and effectively calculate the spectral sum in terms of the two-spinon excitations, allowing us to calculate the one-body function both in terms of the two-spinon excitation and Fredholm determinant, drawing a comparison between those two. Finally, we inspect how the correlator scales with the number of particles in the system. To conclude, we provide our final observations and remarks in Sec. \ref{sec.conclusions}. Then, we show how the correlator can be formulated in terms of Fredholm determinants in App. \ref{sec.Fred}, which we use to compare the results from the spectral sum approximation. Finally, in App. \ref{app.numerical} the reader can find details on numerical procedures used to evaluate rapidities distributions and Fredholm determinants.

\section{The theory}\label{sec.theory}
We consider an interacting quantum field theory in $d=1+1$ with Hamiltonian given by 
 \begin{equation}
     \Ham = \int dx\, \del_x \Psi^\dagger(x) \del_x \Psi(x) + \gamma \int dx \, \Psi^\dagger(x)\Psi^\dagger(x) \Psi(x) \Psi(x) -\eta \int dx \, \Psi^\dagger(x) \Psi(x),
 \end{equation}
  in units where $\hbar = 2m =1$, $\gamma$ is the coupling constant, and $\eta$ is the chemical potential. Here $\Psi(x)$ and $\Psi^\dagger(x)$ are the usual bosonic field operators, which fix the number of particles described by the model 
\begin{equation}
    N = \int dx\, \Psi^\dagger(x) \Psi(x),
\end{equation}
and evolve in time as 
\begin{align}\label{eq.Heisenberg-time-evolution}
    \Psi(x,t) &= e^{i\Ham t} \Psi(x) e^{-i\Ham t}, \\
    \Psi^\dagger(x,t) &= e^{i\Ham t} \Psi^\dagger(x) e^{-i\Ham t}.
\end{align}

The model describes a system composed of interacting particles in one spatial dimension. The theory is also known as the Lieb-Liniger model~\cite{1963_Lieb_PR_130_1,1963_Lieb_PR_130_2}, whose Hamiltonian reads 
\begin{equation}\label{eq.LL_I}
\Ham = -\sum_{j=1}^N \partial_j^2 + 2\gamma  \sum_{j > \ell} \delta (x_j-x_\ell). 
\end{equation}
It is family member of ultralocal theories that can be solved via Bethe ansatz.
Upon imposing periodic boundary conditions over a size $L$, the rapidities are constrained through the Bethe equations 
\begin{equation}\label{eq.Bethe}
e^{i\lambda_j L} = \prod_{j \neq \ell} \frac{\lambda_j - \lambda_\ell + i\gamma}{\lambda_j - \lambda_\ell -i\gamma}, \quad  j,\ell = 1,\dots,N.
\end{equation}

 The solution for the  momenta of the particles is provided by the following set of equations 
\begin{equation}\label{eq.Bethelog}
	\lambda_j+\frac{2}{L} \sum_{\ell =1}^N \tan^{-1}\left(\frac{\lambda_j - \lambda_\ell}{\gamma}\right) = \frac{2\pi}{L}\mathcal{I}_j, \quad  j = 1,\dots,N,
\end{equation}
where $\mathcal{I}_j$ denote the allowed quantum numbers and they present a one-to-one correspondence with the rapidities. 

At zero temperature, the possible quantum numbers obey a Fermi sea structure  
\begin{equation}
	\mathcal{I}_j = j -\frac{N+1}{2} , \quad  j = 1,\dots,N.
\end{equation}

\subsection{Thermal equilibrium}
In the thermodynamically large system, the total density of particles becomes  
\begin{equation}
\frac{N}{L} = \int d\lambda \, \rho(\lambda) ,
\end{equation}
where $\rho(\lambda)$ is the distribution of particles. One may also introduce the distribution of holes $\rho_h(\lambda)$, so that it relates to the latter as  
\begin{equation}\label{eq.roro}
2\pi (\rho(\lambda)+\rho_h(\lambda)) = 1 + \int d\mu \,  K(\lambda,\mu) \rho(\mu).
\end{equation}
Notice here that we have introduced the derivative of the shift between particles that occur in the scattering processes $K(\lambda,\mu):= \partial_\lambda \theta(\lambda), \theta(\lambda):= \tan^{-1}(\lambda/\gamma)$. 

Let the energy of elementary excitations above the equilibrium state be $\varepsilon(\lambda)$, we have  
\begin{equation}\label{eq.15}
\frac{\rho_h(\lambda)}{\rho(\lambda)} = e^{\beta \varepsilon(\lambda)}.
\end{equation}

The partition function is 
\begin{equation}
\mathcal{Z} = \frac{1}{N!} \sum_{n_1,\dots,n_N} \exp{\left( -\beta \sum_{j=1}^N \lambda_{n_j}^2\right)} = \sum_{n_1<n_2<\dots<n_N}\exp{\left( -\beta \sum_{j=1}^N \lambda_{n_j}^2\right)},
\end{equation}
or equivalently 
\begin{equation}
\mathcal{Z} = \sum_{n_{2,1}=1}^\infty\sum_{n_{3,2}=1}^\infty \dots \sum_{n_{N,N-1}=1}^\infty\exp{\left( -\beta \sum_{j=1}^N \lambda_{n_j}^2\right)},
\end{equation}
with $n_{j+1,j} := n_{j+1}-n_{j}$.
In the thermodynamic limit, 
\begin{equation}
\mathcal{Z} \propto \int \mathcal{D}\left\{ \frac{\rho_t(\lambda)}{\rho_p(\lambda)}\right\} 
\delta \left(L \int \rho_p(\lambda) d\lambda -N\right) \exp{\left(S -\beta \sum_{j=1}^N \lambda_{n_j}^2\right)},
\end{equation}
with the entropy is given by 
\begin{equation}\label{eq.entropy}
S = L \int  d\lambda \left[\rho_t(\lambda) \log \rho_t(\lambda)  -\rho(\lambda) \log \rho(\lambda) - \rho_h(\lambda) \log \rho_h(\lambda)  \right],
\end{equation}
where we have defined the total number of vacancies $\rho_t(\lambda) := \rho(\lambda) +\rho_h(\lambda)$. 
This follows from the possible number of arranging $L\rho(\lambda) d\lambda$ particles into $L\rho_t(\lambda) d\lambda$ vacancies, 
\begin{equation}
\frac{\left(L\rho_t(\lambda) d\lambda\right)!}{\left(L\rho(\lambda) d\lambda\right)!\left(L\rho_h(\lambda) d\lambda\right)!} = e^{dS}.
\end{equation}

Such energies can be found by solving the Yang-Yang equation 
\begin{equation}\label{eq.YY}
\varepsilon(\lambda) = \lambda^2 - \eta -\frac{1}{2\pi\beta} \int d\mu \, K(\lambda,\mu) \log\left(1+e^{-\beta \varepsilon(\mu)}\right) .
\end{equation}

\subsubsection{$\gamma\to \infty$}
In the infinite coupling limit, \eqref{eq.YY} becomes 
\begin{equation}
\varepsilon(\lambda) = \lambda^2 - \eta
\end{equation}
and \eqref{eq.roro} together with  \eqref{eq.15} gives 
 \begin{equation}
2\pi \rho (\lambda) = \frac{1}{1+e^{\beta \varepsilon(\lambda)}}, \qquad 2\pi \rho_h (\lambda) = \frac{1}{1+e^{-\beta \varepsilon(\lambda)}}.
\end{equation}

\section{Correlations and form factors}\label{sec.correlations}
In correlated systems, specially in lower dimensions, the inspection of correlation functions provides an insightful 
way to acquire important information on the system. On this note, we specify our primary interest here, which is the 
correlator defined as 
\begin{equation}\label{eq.correlator}
\mathcal{G}(x,t) := \avg{\Psi^\dagger(x,t)\Psi(0,0)},
\end{equation}
where $\avg{\dots}$ denotes quantum-mechanical and thermal average, namely 
\begin{equation}
	\mathcal{G}(x,t) = \frac{\tr{e^{-\beta \mathcal{H}} \Psi^\dagger(x,t)\Psi(0,0)}}{\tr{e^{-\beta \mathcal{H}}}},
\end{equation}
with $\beta = 1/T$ being the inverse temperature, and $\Psi(x,t)$ and $\Psi^\dagger(x,t)$ are bosonic field operators, which can be interpreted as annihilating and creating a particle at position $x$ and time $t$, respectively.

In more practical terms, we have 
\begin{equation}
\mathcal{G}(x,t) = \mathcal{Z}^{-1}\sum_{\bfl} e^{-\beta E_{\bfl}} \frac{\bra{\bfl}\Psi^\dagger(x,t)\Psi(0,0)\ket{\bfl}}{{\braket{\bfl}{\bfl}}},
\end{equation}
where $\mathcal{Z} = \tr{e^{-\beta \mathcal{H}}}$ is the partition function and the set of states $\bfl$ with energy $E_{\bfl}$.

Regarding a thermodynamically large system in equilibrium, we can restrict the sum to a single expectation value \cite{KorepinBOOK}, namely 
\begin{equation}\label{eq.onethermalstate}
\mathcal{G}(x,t) = \frac{\bra{\bfl_T}\Psi^\dagger(x,t)\Psi(0,0)\ket{\bfl_T}}{{\braket{\bfl_T}{\bfl_T}}}.
\end{equation} 
The corresponding Bethe states are quantum mechanically complete, meaning that the correlator can be written as
\begin{equation}\label{eq.G(x,t)2}
\mathcal{G}(x,t) = \sum_{\bfm \in \mathbb{H}_{N-1}} \frac{\bra{\bfl} \Psi^\dagger (x,t) \ket{\bfm} \bra{\bfm} \Psi(0,0) \ket{\bfl}}{\braket{\bfl}{\bfl}\braket{\bfm}{\bfm}},
\end{equation}
where the sum must be performed over all states that are inside the Hilbert space of $N-1$ particles, $\mathbb {H}_{N-1}$. 
The field operators obey the following space and time evolution:  
\begin{equation}
	\Psi^{\dagger}(x,t) = e^{- i \mathcal{H} t + i P x} \Psi^{\dagger}(0) e^{i \mathcal{H} t - i P x}.
\end{equation}
Inserting all these pieces together, we have a spectral representation of the correlator, 
\begin{equation}\label{one-body_cf}
\mathcal{G}(x,t) = \sum_{\bfm \in \mathbb{H}_{N-1}} e^{-i(E_{\bfl}-E_{\bfm})t + i(P_{\bfm}-P_{\bfl})x} |\langle \bfm | \Psi(0) | \bfl \rangle|^2.
\end{equation}
Notice that we have factored out the term $|\langle \bfm | \Psi(0) | \bfl \rangle|^2$, i.e. the space and time independent form factor.

The normalization of the eigenvectors $|\bfl\rangle$ is provided by~\cite{gaudin,korepin} 
\begin{equation}
|| \bfl ||^2 := \braket{\bfl}{\bfl} = \gamma^N \prod_{j>k=1}^N \frac{\lambda_{jk}^2 + \gamma^2}{\lambda_{jk}^2} \det_N \mathbb{M}(\bfl),
	\label{eq.norm}
\end{equation}
where we  have defined for simplicity $\lambda_{jk} := \lambda_j - \lambda_k$. The $\mathbb{M}$ matrix entries are  
\begin{equation}
	\mathbb{M}_{jk}(\bfl) = \delta_{jk} \left(L + \sum_{\ell=1}^N K (\lambda_{j \ell}) \right) - K(\lambda_{jk}), \quad j, k = 1, \dots, N,
	\label{eq.gaudin}
\end{equation}
with 
\begin{equation}\label{eq.kernel}
	K(\lambda)  = \frac{2\gamma}{ \lambda^2 + \gamma^2}.
\end{equation}

The field operator form factors  are given by \cite{one_body_Korepin_Slavnov,1997CMaPh.188..657K,1742-5468-2007-01-P01008}
\begin{equation}
	|\langle \bfm | \Psi(0) | \bfl \rangle|^2 = \gamma^{2N-1} \frac{\prod_{j>k=1}^N \left( \lambda_{jk}^2 + \gamma^2\right)^2}{\prod_{j=1}^N \prod_{k=1}^{N-1} \left(\lambda_j - \mu_k \right)^2} \frac{\det_{N-1}^2 \mathcal{U}(\bfm, \bfl)}{\| \bfm\|^2 \|\bfl\|^2}, \label{ff_LL}
\end{equation}
with the  matrix $\mathcal{U}(\bfm, \bfl)$ elements given by 
\begin{equation}\label{eq.U}
	\mathcal{U}_{jk}(\bfm, \bfl) = \delta_{jk} \frac{\mathcal{V}_j^+ - \mathcal{V}_j^-}{i} + \frac{\prod_{\ell=1}^{N-1}(\mu_\ell - \lambda_j)}{\prod_{\ell \neq j}^N (\lambda_\ell - \lambda_j)} \left[ K(\lambda_{jk}) - K(\lambda_{Nk}) \right], \quad j, k = 1, \dots, N-1,
\end{equation}
where
\begin{equation}
	\mathcal{V}_j^{\pm} = \frac{\prod_{\ell=1}^{N-1} (\mu_\ell - \lambda_j \pm i\gamma)}{\prod_{\ell=1}^N (\lambda_\ell - \lambda_j \pm i\gamma)}.
	\label{eq.Vj}
\end{equation}

\subsection{Strongly repulsive case}
Now we are interested in analysing the form factors \eqref{ff_LL} in the strongly repulsive regime \cite{Panfil_2021}. 
From \eqref{eq.kernel}, $K(\lambda) = 2/\gamma + \mathcal{O}\left(\gamma^{-3}\right)$. Thus,  
\begin{equation}
\left[K(\lambda_{j}-\lambda_{k})-K(\lambda_{N}-\lambda_{k})\right]\sim \mathcal{O}\left(\gamma^{-3}\right).
\end{equation}
Consequently, the leading contribution to the asymptotic behaviour of the matrix elements $\mathcal{U}(\bfm, \bfl)$ comes from $\mathcal{V}_j^{\pm}$, i.e. ~\eqref{eq.Vj}. We find $\mathcal{V}_j^\pm =  \mp i/\gamma + \mathcal{O}\left(\gamma^{-2} \right)$ and therefore
\begin{equation}
\left(\det_{N-1} \mathcal{U}(\bfm, \bfl)\right)^2 = \left(\frac{2}{\gamma}\right)^{2(N-1)} \left( 1 + \mathcal{O}(1/\gamma)\right).
\end{equation}
From \eqref{eq.norm}, we have that \eqref{eq.gaudin} reads
\begin{equation}
\mathbb{M}_{ij}(\bfl_N) = \delta_{ij}\left(L+\frac{2(N-1)}{\gamma}\right)-\frac{2}{\gamma} + \mathcal{O}(1/\gamma^3).
\end{equation}
We observe that the diagonal terms provide the leading contribution to the determinant, so that 
\begin{equation}
\det_{N}\mathbb{M}(\bfl_N) =  \left(L+\frac{2(N-1)}{\gamma}\right)^{N}\times \left(1 + \mathcal{O}(1/\gamma^2) \right) = L^{N} \left(1 + \mathcal{O}(1/\gamma) \right).
\end{equation}
Thence, 
\begin{equation}
\| \bfl_N\|^{2} =  L^{N} \gamma^{N^2}\prod_{j>k=1}^{N}\lambda_{jk}^{-2} \times \left( 1 + \mathcal{O}(1/\gamma) \right),  \quad \| \bfm_{N-1}\|^{2} =(L\gamma)^{N-1}\prod_{j>k=1}^{N-1}\frac{\mu_{jk}^{2}+\gamma^{2}}{\mu_{jk}^{2}}.
\end{equation}
Finally, we have that the form factors in the strong coupling regime yields  
\begin{equation}
 |\langle \bfm | \Psi(0) | \bfl \rangle|^2 =  \left(\frac{2}{L}\right)^{2N-1}\frac{\prod_{j>k=1}^{N}\lambda_{jk}^{2}\prod_{j>k=1}^{N-1}\mu_{jk}^{2}}{\prod_{j=1}^{N}\prod_{k=1}^{N-1}(\lambda_{j}-\mu_{k})^{2}}.
\end{equation}

\section{The two-spinon program}\label{sec.2sp}
Now let us describe how to organize the form factors in terms of two-spinon excitation. The basic idea of the two-spinon is to consider two type II excitations on top of the $(N+1)$-particle state, which we denote by $\bfbm$. Differently from the zero-temperature system, where the holes are punched over the Fermi-sea structure of rapidities distribution, here we employ the two-spinon procedure over a representative thermal state, whose rapidities distribution will be discussed in Sec. \ref{sec.discretisation}.

Following the two-spinon procedure to organize the form factors in terms of two spinon excitations dressed with $m$ particle-hole pairs from \cite{Panfil_2021}, we have 
\begin{align}
	|\langle \bfm | \Psi(0) | \bfl \rangle|^2 =& \, \Omega(L, N) \times \prod_{\ell=1}^{m+2} \frac{\prod_{j=1}^N (\lambda_j -  h_\ell)^2}{\Dprod_{j=1}^{N+1} (\bar{\mu}_j - h_\ell)^2} \prod_{\ell=3}^{m+2} \frac{\prod_{j=1}^{N+1} (\bar{\mu}_j -  p_\ell)^2}{\prod_{j=1}^N (\lambda_j -  p_\ell)^2}\nonumber \\
	&\times \frac{\prod_{j > \ell=1}^{m+2} (h_j -  h_\ell)^2  \prod_{j> \ell=3}^{m+2} (p_j - p_\ell)^2 }{\prod_{j=1}^{m+2} \prod_{\ell=3}^{m+2} (h_j - p_\ell)^2},\label{ff_2sp_mph}
\end{align}
where we introduce the dot product
\begin{equation} \label{d_product}
	\Dprod_{j} \; (\bar{\mu}_j- \bar{\mu}_\ell) := \prod_{\substack{j \neq \ell}} (\bar{\mu}_j -\bar{\mu}_\ell),
\end{equation}
excluding potential zeros and the prefactor is given by 
\begin{equation}
	\Omega(N, L) := \left(\frac{2}{L}\right)^{2N-1} \frac{\prod_{j > k =1}^N (\lambda_j - \lambda_k)^2  \prod_{j > k =1}^{N+1} (\bar{\mu}_j - \bar{\mu}_k)^2}{\prod_{j=1}^N \prod_{k=1}^{N+1} \left(\lambda_j - \bar{\mu}_k \right)^2} .
\label{OmegaNL}
\end{equation}

\subsection{Decreasing form factors}
Let us consider $m$-ph on top of 2sp form factor
\begin{align}
	F_m &= |\langle \bfm | \Psi(0) | \bfl \rangle|^2 = \Omega(L, N) \times \prod_{\ell=1}^{m+2} \frac{\prod_{j=1}^N (\lambda_j -  h_\ell)^2}{\Dprod_{j=1}^{N+1} (\bar{\mu}_j - h_\ell)^2} \prod_{\ell=3}^{m+2} \frac{\prod_{j=1}^{N+1} (\bar{\mu}_j -  p_\ell)^2}{\prod_{j=1}^N (\lambda_j -  p_\ell)^2}\nonumber \\
	&\times \frac{\prod_{j > \ell=1}^{m+2} (h_j -  h_\ell)^2  \prod_{j> \ell=3}^{m+2} (p_j - p_\ell)^2 }{\prod_{j=1}^{m+2} \prod_{\ell=3}^{m+2} (h_j - p_\ell)^2},
\end{align}
and $(m+1)$-ph on top of 2sp form factor
\begin{align}
	F_{m+1} &= |\langle \bfm | \Psi(0) | \bfl \rangle|^2 = \Omega(L, N) \times \prod_{\ell=1}^{m+3} \frac{\prod_{j=1}^N (\lambda_j -  h_\ell)^2}{\Dprod_{j=1}^{N+1} (\bar{\mu}_j - h_\ell)^2} \prod_{\ell=3}^{m+3} \frac{\prod_{j=1}^{N+1} (\bar{\mu}_j -  p_\ell)^2}{\prod_{j=1}^N (\lambda_j -  p_\ell)^2}\nonumber \\
	&\times \frac{\prod_{j > \ell=1}^{m+3} (h_j -  h_\ell)^2  \prod_{j> \ell=3}^{m+3} (p_j - p_\ell)^2 }{\prod_{j=1}^{m+3} \prod_{\ell=3}^{m+3} (h_j - p_\ell)^2}.
\end{align}
We rewrite it as 
\begin{align}
	F_{m+1} &= \Omega(L, N)  \times \frac{\prod_{j=1}^N (\lambda_j -  h_{m+3})^2}{\Dprod_{j=1}^{N+1} (\bar{\mu}_j - h_{m+3})^2} \prod_{\ell=1}^{m+2} \frac{\prod_{j=1}^N (\lambda_j -  h_\ell)^2}{\Dprod_{j=1}^{N+1} (\bar{\mu}_j - h_\ell)^2} \\
            &\times \frac{\prod_{j=1}^{N+1} (\bar{\mu}_j -  p_{m+3})^2}{\prod_{j=1}^N (\lambda_j -  p_{m+3})^2} \prod_{\ell=3}^{m+2} \frac{\prod_{j=1}^{N+1} (\bar{\mu}_j -  p_\ell)^2}{\prod_{j=1}^N (\lambda_j -  p_\ell)^2}\nonumber \\
	&\times  \frac{\prod_{j=1}^{m+2} (h_j -  h_{m+3})^2  \prod_{j=3}^{m+2} (p_j -  p_{m+3})^2}{\prod_{j=1}^{m+2}  (h_j - p_{m+3})^2 \prod_{j=3}^{m+3} (h_{m+3} - p_j)^2}  \frac{\prod_{j > \ell=1}^{m+2} (h_j -  h_\ell)^2  \prod_{j> \ell=3}^{m+2} (p_j - p_\ell)^2 }{\prod_{j=1}^{m+2} \prod_{\ell=3}^{m+2} (h_j - p_\ell)^2},
\end{align}
so that the ratio is 
\begin{equation}
    \frac{F_{m+1}}{F_{m}} = \frac{\prod_{j=1}^N (\lambda_j -  h_{m+3})^2  \prod_{j=1}^{N+1} (\bar{\mu}_j -  p_{m+3})^2}{\Dprod_{j=1}^{N+1} (\bar{\mu}_j - h_{m+3})^2 \prod_{j=1}^N (\lambda_j -  p_{m+3})^2}
     \frac{\prod_{j=1}^{m+2} (h_j -  h_{m+3})^2  \prod_{j=3}^{m+2} (p_j -  p_{m+3})^2}{\prod_{j=1}^{m+2}  (h_j - p_{m+3})^2 \prod_{\ell=3}^{m+3} (h_{m+3} - p_\ell)^2}.
\end{equation}

Now consider the first ratio taking into consideration our symmetric discretisation $\lambda_j = (\bar{\mu}_j + \bar{\mu}_{j+1} )/2$:
\begin{equation}
    \frac{\prod_{j=1}^N ( (\bar{\mu}_j + \bar{\mu}_{j+1} )/2 -  h_{m+3})^2 }{\Dprod_{j=1}^{N+1} (\bar{\mu}_j - h_{m+3})^2} = 
     \left( \frac{\prod_{j=1}^N ((\bar{\mu}_j + \bar{\mu}_{j+1} )/2 -  h_{m+3}) }{\Dprod_{j=1}^{N+1} (\bar{\mu}_j - h_{m+3})} \right)^2.
\end{equation}
Proving that 
\begin{equation}
    \Bigg| \frac{\prod_{j=1}^N ((\bar{\mu}_j + \bar{\mu}_{j+1} )/2 -  h_{m+3}) }{\Dprod_{j=1}^{N+1} (\bar{\mu}_j - h_{m+3})} \Bigg| <1
\end{equation}
is sufficient, so we proceed with that. 
Also, let us fix the position of the hole $h_{m+3} = \bar{\mu}_h$. Thence, 
\begin{equation}
    \begin{aligned}
    \frac{\prod_{j=1}^N ((\bar{\mu}_j + \bar{\mu}_{j+1} )/2 -  h_{m+3}) }{\Dprod_{j=1}^{N+1} (\bar{\mu}_j - h_{m+3})} 
    &=  \frac{\prod_{j=1}^N \frac{1}{2}  (\bar{\mu}_j + \bar{\mu}_{j+1} -  2h_{m+3}) }{\Dprod_{j=1}^{N+1} (\bar{\mu}_j - h_{m+3})} \\
    &= \left( \frac{1}{2} \right)^N  \frac{\prod_{j=1}^{h-1} (\bar{\mu}_j + \bar{\mu}_{j+1} -  2h_{m+3}) \prod_{j=h+1}^{N} (\bar{\mu}_j + \bar{\mu}_{j+1} -  2h_{m+3})}{\prod_{j=1}^{h-1} (\bar{\mu}_j - h_{m+3}) \prod_{j=h+1}^{N} (\bar{\mu}_j - h_{m+3}) } 
    \frac{(\bar{\mu}_{h+1} -   \bar{\mu}_h)}{(\bar{\mu}_{N+1} -  \bar{\mu}_h)} \\ 
    &= \frac{(\bar{\mu}_{h+1} -   \bar{\mu}_h)}{(\bar{\mu}_{N+1} -  \bar{\mu}_h)}  \prod_{j=1, j\neq h}^{N}  \, \frac{1}{2} \left(1 + \frac{\bar{\mu}_{j+1} -  \bar{\mu}_h}{\bar{\mu}_j -  \bar{\mu}_h} \right).
    \end{aligned}
\end{equation}
We observe that 
\begin{equation}
    |\bar{\mu}_{h+1} -   \bar{\mu}_h| \leq |\bar{\mu}_{N+1} -  \bar{\mu}_h|.
\end{equation}
Also, by making use of the difference between adjacent rapidities $\bar{\mu}_{j+1} - \bar{\mu}_{j} = \mathcal{O}\left(1/L\right)$, we have  
\begin{equation}
    \begin{aligned}
        \frac{1}{2} \left( 1 + \frac{\bar{\mu}_{j+1} -  \bar{\mu}_h}{\bar{\mu}_j -  \bar{\mu}_h} \right) &= 
    \frac{1}{2} \left( 1 + \frac{\bar{\mu}_{j} -  \bar{\mu}_h + \mathcal{O}\left(1/L\right)}{\bar{\mu}_j -  \bar{\mu}_h} \right) \\
    &= \left( 1 + \frac{\mathcal{O}\left(1/L\right)}{2(\bar{\mu}_j -  \bar{\mu}_h)} \right) \\
    &= \left( 1 + \frac{\mathcal{O}\left(1/L\right)}{2(j-h)\mathcal{O}\left(1/L\right)} \right) \\
    &= \left( 1 + \frac{1}{2(j-h)} \right),
    \end{aligned}
\end{equation}
so that we are left to consider the product 
\begin{equation}
    \prod_{j=1, j\neq h}^{N}  \left( 1 + \frac{1}{2(j-h)} \right).
\end{equation}
Notice that $j-h = \pm 1, \pm 2, \dots, \pm N$, thus 
\begin{equation}
    1+\frac{1}{2(j-h)} = 1 \pm \frac{1}{2}, 1 \pm \frac{1}{4}, \dots, 1 \pm \frac{1}{2N}.
\end{equation}
Observing that 
\begin{equation}
    \left( 1+\frac{1}{2(j-h)} \right)   \left( 1-\frac{1}{2(j-h)} \right) =  1-\frac{1}{4(j-h)^2} < 1, 
\end{equation}
we conclude that 
\begin{equation}
     \prod_{j=1, j\neq h}^{N} \, \frac{1}{2} \left(1 + \frac{\bar{\mu}_{j+1} -  \bar{\mu}_h}{\bar{\mu}_j -  \bar{\mu}_h} \right) <1.
\end{equation}

Therefore, we have proven the inequality 
\begin{equation}
     \frac{\prod_{j=1}^N (\lambda_j -  h_{m+3})^2  \prod_{j=1}^{N+1} (\bar{\mu}_j -  p_{m+3})^2}{\Dprod_{j=1}^{N+1} (\bar{\mu}_j - h_{m+3})^2 \prod_{j=1}^N (\lambda_j -  p_{m+3})^2} < 1.
\end{equation}

Finally, we are only left to consider 
\begin{equation}
    \frac{F_{m+1}}{F_{m}} = 
    \frac{\prod_{j=1}^{m+2} (h_j -  h_{m+3})^2  \prod_{j=3}^{m+2} (p_j -  p_{m+3})^2}{\prod_{j=1}^{m+2}  (h_j - p_{m+3})^2 \prod_{j=3}^{m+3} (p_j - h_{m+3})^2}.
\end{equation}
We can readily realize that 
\begin{equation}
\begin{aligned}
       & \prod_{j=1}^{m+2} (h_j -  p_{m+3})^2 > \prod_{j=1}^{m+2} (h_j -  h_{m+3})^2, \\
        & \prod_{j=3}^{m+2} (p_j -  p_{m+3})^2 < \prod_{j=3}^{m+2} (p_j -  h_{m+3})^2, 
\end{aligned}
\end{equation}
as $|p_{m+3}| > |k_F|$ and $|h_{m+3}| < |k_F|$.

Therefore we conclude that $F_{m+1} < F_m$. This means that whenever we consider a further particle-hole excitation, the corresponding form factor is smaller than the one from one less ph excitation, providing a smaller contribution to the correlator.

\subsection{Discretisation}\label{sec.discretisation}
Now we want to effectively evaluate the two-spinon correlator,
\begin{equation}\label{eq.2sp}
	\mathcal{G}_{2{\rm sp}}(x,t) = \sum_{\ket{\bfm} \in \mathbb{H}_{2sp}} e^{i \omega_{2{\rm sp}} t - i k_{2{\rm sp}} x} F_{\rm 2sp},
\end{equation}
where $\ket{\bfm}$ is the two-spinon excited state constructed from the $(N+1)$-particle elementary state, and $\ket{\bfl}$ is the $N$-particle elementary state, where 
\begin{equation}
        \omega_{2{\rm sp}} = 2k_F^2 - h_1^2 - h_2^2, \quad  k_{2{\rm sp}} = - h_1 - h_2,
\end{equation}
are the energy and momentum of the two-spinon state, respectively. 
For $m=0$, meaning we consider only two-spinon excitation, we have that the form factors read 
\begin{equation}
	F_{\rm 2sp} = \Omega(N,L) \times \prod_{\ell=1}^{2} \frac{\prod_{j=1}^N (\lambda_j -  h_\ell)^2}{\Dprod_{j=1}^{N+1} (\bar{\mu}_j - h_\ell)^2} \prod_{j > \ell=1}^{2} (h_j -  h_\ell)^2, 
\end{equation}
with the prefactor given by \footnote{When numerically evaluating the prefactor, the products of differences between rapidities can be very small, leading to numerical error. The trick to evaluate it is to rewrite the products as 
\begin{equation}
\prod_{j > k =1}^N (\lambda_j - \lambda_k)^2 =  \exp{\left(\sum_{j\neq k}\log |\lambda_j-\lambda_k|\right)} \overset{L\to \infty}{\Longrightarrow} \exp{\left(\int d\lambda_1 d\lambda_2 \rho(\lambda_1)\rho(\lambda_2)  \log |\lambda_1-\lambda_2| \right)}.
\end{equation}
}
\begin{equation}
	\Omega(N, L) = \left(\frac{2}{L}\right)^{2N-1} \frac{\prod_{j > k =1}^N (\lambda_j - \lambda_k)^2  \prod_{j > k =1}^{N+1} (\bar{\mu}_j - \bar{\mu}_k)^2}{\prod_{j=1}^N \prod_{k=1}^{N+1} \left(\lambda_j - \bar{\mu}_k \right)^2}.
\end{equation}

To practically make use of the developed framework, we introduce a counting function in order to discretise the rapidities,  
\begin{equation}\label{eq.discretisation}
    j  + \frac{1}{2} = \frac{L}{2\pi} \int_{-\infty}^{\mu_j} \rho(\lambda) d\lambda.
\end{equation}
In order to have a symmetric distribution of rapidities, we consider the following shift  
\begin{equation}
\begin{aligned}
\mu_j  &\to \frac{1}{2}\left(\mu_j - \mu_{N+2-j} \right), \quad j=1,\dots,N+1, \\ 
\lambda_j &\to \frac{1}{2}\left(\mu_j + \mu_{j+1}  \right), \quad j=1,\dots,N.
\end{aligned}
\end{equation}
It is easy to check that as $\beta \to \infty$, $\{\bfl \}\big|_{T>0} \to \{\bfl \}\big|_{T=0}$, where $\{\bfl \}\big|_{T=0}, 
 \lambda_j = 2\pi(j- (N+1)/2)/L$, namely our counting function provides the ground-state rapidities discretisation, or Fermi sea structure, at zero temperature.

\begin{figure}[h!]
\centering
\begin{subfigure}{0.45\textwidth}
    \includegraphics[width=\textwidth]{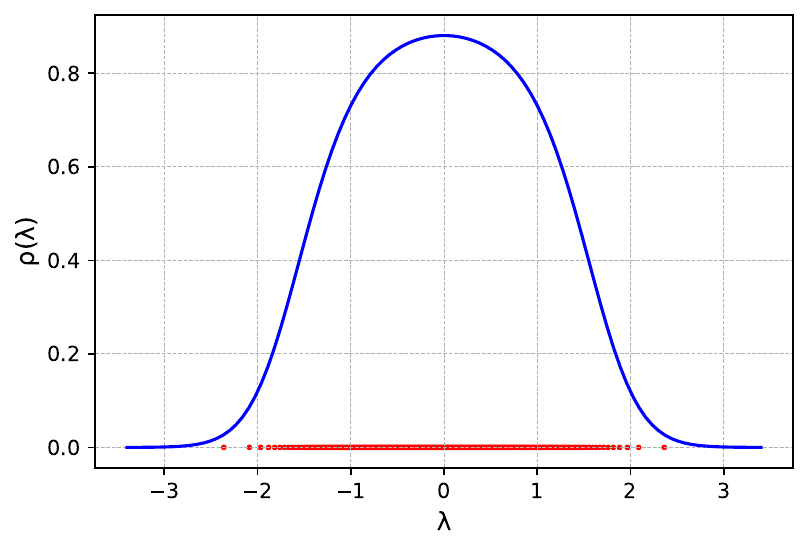}
\end{subfigure}
\hfill
\begin{subfigure}{0.5\textwidth}
    \includegraphics[width=\textwidth]{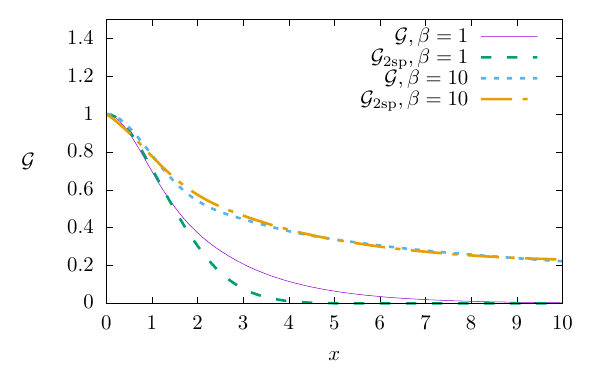}
\end{subfigure}
\caption{On the left-hand side we have the thermal ($\beta=1$) rapidities distribution in red, following the discretisation \eqref{eq.discretisation} for $N=200$ particles measured by the distribution $\rho(\lambda)$ in blue. On the right-hand size, the correlator \eqref{eq.correlator} evaluated at $t=0$ both in terms of the Fredholm determinant representation from \eqref{eq.Fredholm} and by means of the two-spinon program from \eqref{eq.2sp} with two particle-hole excitations on top, for two inverse temperatures $\beta=1,10$.}
\label{fig.rhovsg2sp_t0}
\end{figure}

Finally, in fig. \ref{fig.rhovsg2sp_t0}  we display the resulting correlators at different temperatures. Although we have demonstrated that the contribution to calculating the correlator decreases as one adds further particle-hole excitations on top of the two spinon, we still need to consider a few of them in order to get good convergence with the exact one. To that end, in our numerical evaluation we consider two particle-hole excitations (2ph) on top of the two-spinon program.  
We clearly see that the importance of the two-spinon excitation increases at lower temperatures. This can be explained by the fact that at lower temperatures quantum fluctuations become more important for the dynamics of the system. In contrast, when one increases the temperature, thermal fluctuations overcome quantum effects. Therefore, the two-spinon contribution is obviously a low temperature effect. It is worth mentioning that we restrict ourselves to a maximum inverse temperature of $\beta=10$ due to the fact that higher orders of magnitude originate almost identical, overlapping curves.

\subsection{Scaling}
At this point, we argue that we need to rescale the correlator in order to have a finite result, as we should expect that \eqref{eq.2sp} decays as $e^{-N}$. Therefore, taking into consideration that there are $e^S$ arbitrary states that one could consider in \eqref{eq.onethermalstate}, where $S$ is the entropy given by \eqref{eq.entropy}, we make the shift $\mathcal{G}_{\rm 2sp} \to e^S \mathcal{G}_{\rm 2sp}$. With this motivation, let us inspect how the form factors scale with the number of particles $N$. 
We conjecture that the prefactor \eqref{OmegaNL}  has an implicit exponential decay in terms of $N$, scaling as  
\begin{equation}
    \Omega \sim  e^{-\xi(\beta) N}.
\end{equation}

\begin{figure}[h!]
\centering
\begin{subfigure}{0.45\textwidth}
    \includegraphics[width=\textwidth]{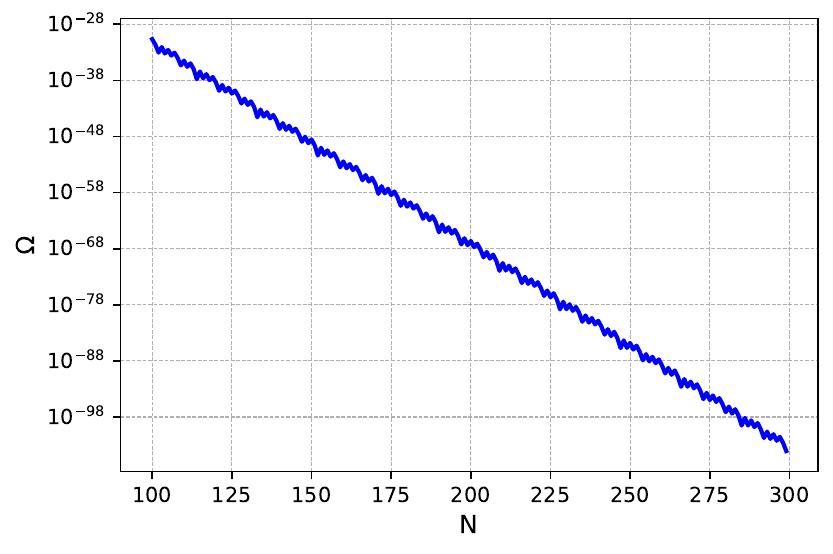}
\end{subfigure}
\hfill
\begin{subfigure}{0.45\textwidth}
    \includegraphics[width=\textwidth]{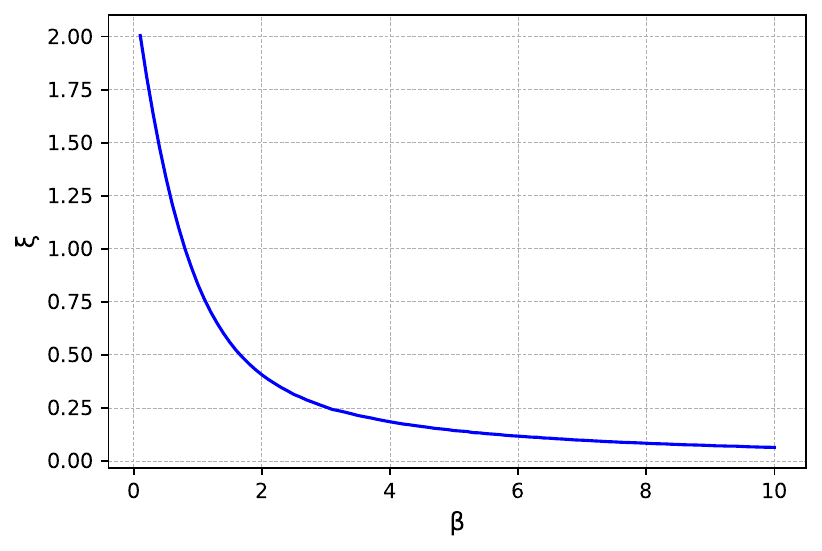}
\end{subfigure}
\caption{Prefactor scaling as $\Omega \sim  e^{-\xi(\beta) N}$. On the left-hand side, we see the relation between the prefactor and the number of particles in log scale displaying its exponential decay, while on the right-hand side the scaling parameter $\xi$ as function of the inverse temperature.}
\label{fig.scale}
\end{figure}

In order to verify that, we numerically evaluate the prefactor for different number of particles, ranging from 100 to 300, making use of the rapidities distribution \eqref{eq.discretisation}. The result is shown in fig. \ref{fig.scale}. On the left-hand side, we observe a linear decay of the prefactor in terms of the number of particles in log scale, corroborating the exponential decay. On the right-hand side we observe the temperature-dependent parameter that dictates the slope of the exponential decay.  Additionally, we can observe that $\xi(\beta) \sim  2e^{-\beta}$, thus $\Omega \sim \exp \left(-2e^{-\beta} N\right)$.

\begin{figure}[h!]
\centering
\begin{subfigure}{0.45\textwidth}
    \includegraphics[width=\textwidth]{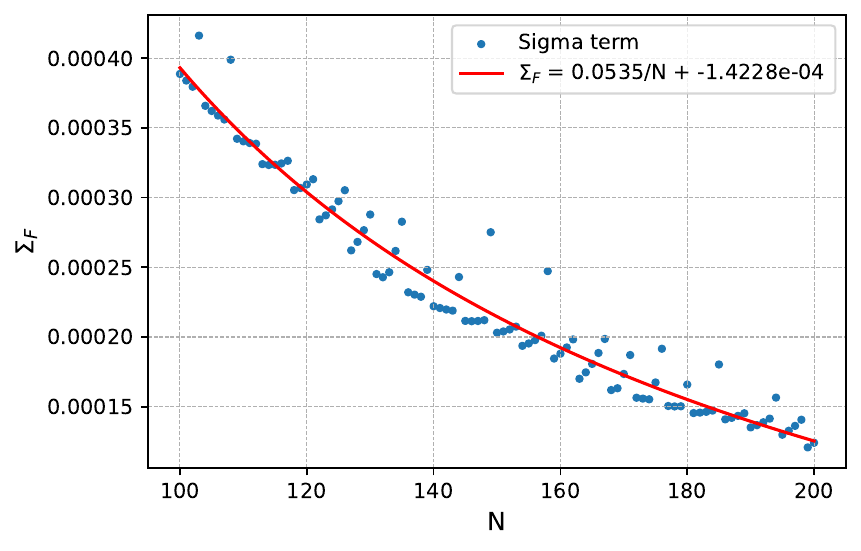}
\end{subfigure}
\hfill
\begin{subfigure}{0.45\textwidth}
    \includegraphics[width=\textwidth]{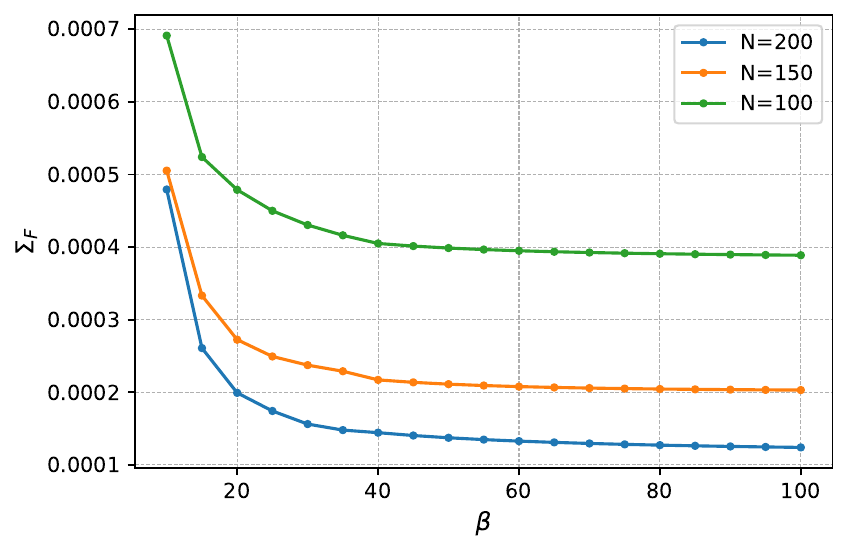}
\end{subfigure}
\caption{On the left-hand side we plot the sigma term \eqref{eq.sum_ff} as a function of the number of particles $N$ for the inverse temperature $\beta=100$. On the right-hand side, we plot the sigma term as function of the inverse temperature $\beta$ for $N=100,150,200$ particles.}
\label{fig.sum_ff}
\end{figure}

Therefore, with the prefactor scaling we observe that $\mathcal{G}_{\rm 2sp} \sim e^S e^{-\xi(\beta) N}   \sum_{\ket{\bfm} \in \mathbb{H}_{2sp}}\tilde{F}_{\rm 2sp}  $ , where  $\tilde{F}_{\rm 2sp} := F_{\rm 2sp}/\Omega$ are the rescaled form factors. Namely, 
\begin{equation}
  \tilde{F}_{\rm 2sp} =  \prod_{\ell=1}^{2} \frac{\prod_{j=1}^N (\lambda_j -  h_\ell)^2}{\Dprod_{j=1}^{N+1} (\bar{\mu}_j - h_\ell)^2} \prod_{j > \ell=1}^{2} (h_j -  h_\ell)^2.
\end{equation}
Breaking the terms down:
\begin{equation}
         \begin{aligned}
             \prod_{j > \ell=1}^{2} (h_j -  h_\ell)^2 \sim \mathcal{O} \left(\frac{h_1-h_2}{L} \right)^2,
         \end{aligned}
     \end{equation}
     \begin{equation}
         \begin{aligned}
          \prod_{j=1,j\neq h_\ell}^{N+1} (\bar{\mu}_j - h_\ell)^2 \sim  \prod_{j=1,j\neq h_\ell}^{N+1} \mathcal{O} \left(\frac{j-h_\ell}{L} \right)^2 \sim \frac{1}{L^{2N}}  \prod_{j=1,j\neq h_\ell}^{N+1} (j-h_\ell)^2 = \frac{1}{L^{2N}} \Gamma(h_\ell)^2 \Gamma(N+2 -h_\ell)^2,
         \end{aligned}
     \end{equation}
 \begin{equation}
         \begin{aligned}
          \prod_{j=1}^{N} (\lambda_j - h_\ell)^2 \sim   \prod_{j=1}^{N}  \mathcal{O} \left(\frac{j-h_\ell}{L} + \frac{1}{2L}\right)^2 \sim \frac{1}{L^{2N}} \prod_{j=1}^{N} \left(j-h_\ell + \frac{1}{2}\right)^2 = \frac{1}{L^{2N}}  \frac{\Gamma \left( N + 3/2 -h_\ell \right)^2}{\Gamma \left( 3/2 -h_\ell \right)^2},
         \end{aligned}
     \end{equation}
     we have that 
     \begin{equation}
  \tilde{F}_{\rm 2sp}\sim   \frac{\Gamma \left( N + 3/2 -h_1 \right)^2}{ \Gamma(h_1)^2 \Gamma(N+2 -h_1)^2\Gamma \left( 3/2 -h_1 \right)^2}   \frac{\Gamma \left( N + 3/2 -h_2 \right)^2}{\Gamma(h_2)^2 \Gamma(N+2 -h_2)^2 \Gamma \left( 3/2 -h_2 \right)^2}  \left(\frac{h_1-h_2}{L} \right)^2.
\end{equation}
Let us consider $h_1,h_2 \ll N$.   Making use of the Lanczos approximation for the large argument limit of the gamma function, i.e. 
$\Gamma(z) \approx \sqrt{2 \pi} z^{z-1/2} e^{-z}, z\to \infty$, the rescaled form factors reduce to 
 \begin{equation}
  \begin{aligned}
      \tilde{F}_{\rm 2sp} & \sim \left(\frac{h_1-h_2}{L} \right)^2 \prod_{\ell=1}^2  \left(\frac{\left(N+3/2-h_\ell \right)^{N+1-h_\ell} e^{N+3/2-h_\ell }}{\left(N+2-h_\ell \right)^{N+3/2-h_\ell} e^{N+2-h_\ell }}\right)^2  \frac{1}{ \Gamma(h_\ell)^2 \Gamma \left( 3/2 -h_\ell \right)^2}  \\
  &\sim   \left(\frac{h_1-h_2}{L}\right)^2 \frac{N^{-2} e^{-1/2}  }{\prod_\ell \Gamma(h_\ell)^2 \Gamma \left( 3/2 -h_\ell \right)^2}.
  \end{aligned}
\end{equation}
In the opposite scenario, i.e. $h_1,h_2 \sim N$, we have 
 \begin{equation}
  \begin{aligned}
      \tilde{F}_{\rm 2sp} & \sim \left(\frac{h_1-h_2}{L} \right)^2 \prod_\ell \frac{\Gamma \left( N + 3/2 -h_\ell \right)^2}{\Gamma(h_\ell)^2 \Gamma \left( N + 2 -h_\ell \right)^2}\pi^{-2} \sin^2 \pi \left( \frac{3}{2} -h_\ell\right) \Gamma \left(h_\ell - \frac{1}{2}\right)^2,
  \end{aligned}
\end{equation}
where we have used the gamma reflection formula in order to avoid the poles at negative integers. 
Now, summing over the $N(N+1)/2$ form factors, we argue that their contribution balance out, up to a small correction. 
We observe this fact by numerically evaluating the sum of the rescaled form factors, 
\begin{equation}\label{eq.sum_ff}
    \Sigma_F :=  \sum_{\ket{\bfm} \in \mathbb{H}_{2sp}} \tilde{F}_{\rm 2sp},
\end{equation}
in terms of the number of particles. We call it the sigma term. Let us inspect how it corresponds to $N$. Empirically from fig. \ref{fig.sum_ff}, we numerically infer the scaling factor $\Sigma_F \sim a/\beta + b$, with parameters given by $\{(a,b)\} = \{(0.0031, 0.00034), (0.00301, 0.00015562), (0.0034, 7.092{\rm e}-05)\}$ for the respective increasing number of particles $N=100,150,200$. We readily recognize $a\sim 1/N$ from the left-hand side.
 Consequently, we have that the two-spinon correlator scales as $\mathcal{G}_{\rm 2sp} \sim e^S e^{-\xi(\beta) N}  /N\beta$.

\section{Conclusions}\label{sec.conclusions}
In this work we have shown how to efficiently calculate expectation values of the field-field operator of the Lieb-Liniger model in its infinitely repulsive regime, also known as Tonks-Girardeau gas, in terms of intermediary states at finite temperatures. By means of organizing the form factors in terms of two-spinon excitations, we have shown that they provide a reliable basis for evaluating the spectral sum of the correlator at low enough temperatures. We analytically showed that by adding particle-hole excitations on top of the two-spinon gradually provides gradually smaller contributions. Finally, we have numerically evaluated both the Fredholm determinant of the theory as well as the spectral sum in terms of the two-spinon excitation and showed that the the results converge as the temperature decreases. This highlights two scenarios: the first one where we can effectively calculate the one-body correlation function of the Tonks-Girardeau gas at low enough temperatures by means of the two-spinon program; and another one that, in contrast, is characterized by high temperatures and further particle-hole excitations on top of two-spinon is shown to be necessary for good convergence between the full spectral sum and the two-spinon program. We hope our work inspires further investigation on the intermediate states significance to integrable models, as it seems to provide a promising direction to acquire invaluable insights on correlation aspects of interacting quantum systems in reduced dimensionality.

\section{Acknowledgements}
We acknowledge Mi{\l}osz Panfil and Oleksandr Gamayun for fruitful discussions. 
This research is part of the project No. 2021/43/P/ST2/02904 co-funded by the
National Science Centre and the European Union Framework Programme for Research
and Innovation Horizon 2020 under the Marie Skłodowska-Curie grant agreement No.
945339. For the purpose of Open Access, the authors have applied a CC-BY public
copyright licence to any Author Accepted Manuscript (AAM) version arising from this
submission.

\appendix 

\section{Fredholm determinant representation}\label{sec.Fred}
In this section we derive the Fredholm determinant representation of the correlator \eqref{eq.G(x,t)2}. For that, we define the form factors as 
\begin{equation}
    \mathcal{F} := \bra{\bfl} \Psi^\dagger(x,t) \ket{\bfm},
\end{equation}
with the corresponding eigenvectors  
\begin{equation}
\begin{aligned}
        \bra{\bfl} &= \frac{1}{\sqrt{N!}} \int \prod_{j=1}^N dx_j \; \varphi_N(x_1,\dots, x_N) \bra{0}\left \{ \bigotimes\limits_{j=1}^N  \Psi^\dagger(x_j) \right \} , \\
            \ket{\bfm} &= \frac{1}{\sqrt{(N-1)!}} \int \prod_{j=1}^{N-1} dx_j \; \varphi^*_N(x_1,\dots, x_N)   \left \{ \bigotimes\limits_{j=1}^N  \Psi(x_j) \right \} \ket{0},
\end{aligned}
\end{equation}
where $\varphi_N$  are the wave functions of the Lieb-Liniger model given by  
\begin{equation}\label{eq.eigenfun}
    \varphi_N = \mathcal{N}^{-1} \exp \left( \sum_{j > \ell} \sgn(x_j-x_\ell)  \right) \sum\limits_{\mathfrak{S} \in \mathcal{P}_N} (-1)^\mathfrak{S} \exp \left( i\sum_{j=1}^N x_j \lambda_\mathfrak{S} \right) \prod_{j>\ell} \left(\lambda_{\mathfrak{S}_{j\ell}} -i\gamma \, \sgn(x_j-x_\ell)   \right),
\end{equation}
where 
\begin{equation}
    \mathcal{N} = N! \prod_{j>\ell} \left(\lambda_{j\ell} + \gamma^2 \right),
\end{equation}
and $\mathcal{P}_N$ is the permutation group of $N$ elements, while $(-1)^\mathfrak{S}$ corresponds to the sign of the permutation. 

With that, the form factors are 
\begin{equation}
    \begin{aligned}
        \mathcal{F} =& \left(\frac{1}{\sqrt{N!}} \int \prod_{j=1}^N dx_j \; \varphi_N(x_1,\dots, x_N) \bra{0} \left \{ \bigotimes\limits_{j=1}^N  \Psi^\dagger(x_j) \right \} \right) \\ 
         &\times \Psi(x,t) \left( \frac{1}{\sqrt{(N-1)!}} \int \prod_{j=1}^{N-1} dx_j \; \varphi^*_N(x_1,\dots, x_N)   \left \{ \bigotimes\limits_{j=1}^N  \Psi(x_j) \right \} \ket{0}\right) 
    \end{aligned}
\end{equation}

Now we specialize at $t=0$. In this case the form factors become (by applying the commutation rules of the bosonic field operators)
\begin{equation}\label{eq.ffB4}
    \mathcal{F}(x) = \sqrt{N-1}  \int \prod_{j=1}^{N-1} dx_j \; \varphi^*_{N+1}(x_1,\dots, x_N,x)   \varphi_N(x_1,\dots, x_N).
\end{equation}

The wave functions \eqref{eq.eigenfun} can be conveniently rewritten as 
\begin{equation}\label{eq.wf_LL}
    \varphi_N(x_1,\dots,x_N|\{\bfl\}) = \frac{1}{\sqrt{N!}} \prod\limits_{j>\ell} {\rm sgn}  (x_j-x_\ell) \sum\limits_{\mathfrak{S} \in \mathcal{P}_N} (-1)^\mathfrak{S}  \exp \left( i \sum_{j=1}^N x_j \lambda_{\mathfrak{S}_j} \right).
\end{equation}
Substituting \eqref{eq.wf_LL} into \eqref{eq.ffB4}: 
\begin{equation}
    \begin{aligned}
        \mathcal{F}(x) =& \frac{1}{N!}  \sum_{\mathfrak{S}_1 \in \mathcal{P}_{N}}  \sum_{\mathfrak{S}_{2} \in \mathcal{P}_{N+1}}  (-1)^{\mathfrak{S}_1 + \mathfrak{S}_2} e^{-ix \lambda_{\mathfrak{S}_{2,N+1}}} \\
        &\times \int \prod_{j=1}^N dx_j \sgn(x-x_j) \exp \left(  -i\sum_{j=1}^N (\mu_{\mathfrak{S}_{2,j}} - \lambda_{\mathfrak{S}_{1,j}}) x_j \right). 
    \end{aligned}
\end{equation}
Here we take a note: the sum is taken over the permutations $\mathfrak{S}_1$ and $\mathfrak{S}_2$ between the sets of rapidities $\{\lambda_j\}, \, j=1, \dots, N$ and $\{\mu_j\}, \, j=1, \dots, N-1$. 

The integral can be evaluated, yielding 
\begin{equation}
    \mathcal{F}(x) = \frac{(2i)^N}{N!} e^{ ix (\Lambda -\mathcal{M}) }  \sum_{\mathfrak{S}_1 \in \mathcal{P}_{N}}  \sum_{\mathfrak{S}_{2} \in \mathcal{P}_{N+1}} (-1)^{\mathfrak{S}_1 + \mathfrak{S}_2} 
    \prod_{j=1}^N \left( \mu_{\mathfrak{S}_{2,j}} - \lambda_{\mathfrak{S}_{1,j}} \right)^{-1},
\end{equation}
where 
\begin{equation}
        \Lambda := \sum_{j=1}^N \lambda_j, \quad  \mathcal{M} := \sum_{j=1}^{N+1} \mu_j.
\end{equation}
By making use of the identity 
\begin{equation}
    \begin{aligned}
        &\frac{1}{N!} \sum_{\mathfrak{S}_1 \in \mathcal{P}_{N}}  \sum_{\mathfrak{S}_{2} \in \mathcal{P}_{N+1}} (-1)^{\mathfrak{S}_1 + \mathfrak{S}_2} 
    \prod_{j=1}^N \left( \mu_{\mathfrak{S}_{2,j}} - \lambda_{\mathfrak{S}_{1,j}} \right)^{-1} \\
    &= (1+\partial_\xi) \det\limits_N \left( \frac{1}{\mu_j-\lambda_\ell} -\frac{\xi}{\mu_{N+1} -\lambda_\ell} \right)\Bigg|_{\xi=0},
    \end{aligned}
\end{equation}
we arrive at 
\begin{equation}
    \mathcal{F}(x) = (2i)^N e^{ix (\Lambda-\mathcal{M})} (1+\partial_\xi) \det\limits_N \left( \frac{1}{\mu_j-\lambda_\ell} -\frac{\xi}{\mu_{N+1} -\lambda_\ell} \right)\Bigg|_{\xi=0}.
\end{equation}
The time dependence of the above expression can be introduced by considering the time evolution of the bosonic field operator from \eqref{eq.Heisenberg-time-evolution},  
\begin{equation}
    \mathcal{F}(x,t) = (2i)^N e^{it(E_\Lambda-E_\mathcal{M} -\eta)} e^{ix (\Lambda-\mathcal{M})} (1+\partial_\xi) \det\limits_N \left(\mathcal{D}_{j,\ell} \right)\Big|_{\xi=0},
\end{equation}
where we define the matrix $\mathcal{D}$ with elements
\begin{equation}
\mathcal{D}_{j\ell}:= \left( \frac{1}{\mu_j-\lambda_\ell} -\frac{\xi}{\mu_{N+1} -\lambda_\ell} \right), \quad  j,\ell=1,\dots,N, 
\end{equation}
and 
\begin{equation}
        E_\Lambda := \sum_{j=1}^N \lambda_j^2, \quad  E_\mathcal{M} := \sum_{j=1}^{N-1} \mu_j^2. 
\end{equation}

Now that we have an expression for the form factors, let us insert it into the correlator \eqref{eq.G(x,t)2} 
\begin{equation}
\begin{aligned}
\mathcal{G}(x,t) &= \frac{1}{L^{2N-1}} \sum_{\bfm \in \mathbb{H}_{N-1}}  \mathcal{F}(x,t) \mathcal{F}^*(0,0) \\
& = \frac{4^{N}}{L^{2N-1}} \sum_{\bfm \in \mathbb{H}_{N-1}}  e^{it(E_\Lambda-E_\mathcal{M} -\eta)} e^{ix (\Lambda-\mathcal{M})} \Bigg| (1+\partial_\xi) \det\limits_N \left(\mathcal{D}_{j,\ell} \right)\Big|_{\xi=0}\Bigg|^2
\end{aligned}
\end{equation}
where we have considered that $\braket{\bfl}{\bfl} \sim \mathcal{O}(L^{N})$ and $\braket{\bfm}{\bfm} \sim \mathcal{O}(L^{N-1})$.
With some algebra it is possible to rewrite the last equality as 
\begin{equation}\label{eq.g.sum}
\begin{aligned}
\mathcal{G}(x,t) &= \frac{(N-1)!}{L} \left(\frac{2}{L}\right)^{2N}  \sum_{\bfm \in \mathbb{H}_{N-1}}  \left[ e^{it(E_\Lambda-E_\mathcal{M} -\eta)} e^{ix (\Lambda-\mathcal{M})}   + \partial_\xi \right] \\
&\times \det  \left( \frac{e^{it\mu_j^2+ix\mu_j}}{(\mu_j-\lambda_\ell)(\mu_j-\lambda_j)} -\xi \frac{e^{it\mu_j^2+ix\mu_j}}{(\mu_j-\lambda_j)}  \frac{e^{it\mu_{N-1}^2+ix\mu_{N-1}}}{(\mu_{N-1}-\lambda_\ell)}    \right)\Bigg|_{\xi=0}.
\end{aligned}
\end{equation}
Now, let us rewrite the sum over all $\bfm$ by individual sums, namely 
\begin{equation}\label{eq.sums}
    \sum_{\bfm \in \mathbb{H}_{N-1}} = \frac{1}{(N-1)!}  \sum_{\mu_1}  \sum_{\mu_2} \cdots  \sum_{\mu_{N-1}}, \quad \mu_j \in \mathbb{Z}, \quad j=1,2,\cdots,N-1. 
\end{equation}
With these considerations, the correlator yields \footnote{For further mathematical details, we refer to \cite{KorepinBOOK,patu_I,patu_II}.}
\begin{equation}\label{eq.correlator_sum}
\begin{aligned}
&\mathcal{G}(x,t) = e^{i \eta t} \left(\frac{1}{2\pi} \Upsilon(x,t) + \del_\xi\right) \\
&\times \det \left(\delta_{j,\ell} \frac{2\Xi(x,t,\lambda_\ell)}{\pi L} e^{it\lambda_\ell^2-ix\lambda_\ell} +e^{-it\lambda_\ell^2 +ix\lambda_\ell} \frac{2(1-\delta_{j,\ell})}{\pi L (\lambda_j-\lambda_\ell)} (\Xi(x,t,\lambda_j) -\Xi(x,t,\lambda_\ell)) -\frac{\xi}{L \pi^2} \Xi(x,t,\lambda_j) \Xi(x,t,\lambda_\ell)\right)\Bigg|_{\xi=0},
\end{aligned}
\end{equation}
with 
\begin{equation}
    \Upsilon(x,t):= \frac{2\pi}{L}\sum_{\mu_{N-1}} e^{it\mu_{N-1}^2-ix\mu_{N-1}}\, \quad \Xi(x,t,\lambda_\ell):= \frac{2\pi}{L} \sum_{\mu_j} \frac{e^{it\mu_j^2-ix\mu_j}}{\mu_j-\lambda_\ell}.
\end{equation}

Now, let us analyse the thermodynamic limit of the summations in \eqref{eq.correlator_sum}. 
The sums are performed over the function $f(\mu_j) = e^{it\mu_j^2-ix\mu_j}$. Firstly, we have
\begin{equation}
    S_0 = \frac{1}{L} \sum_{\mu_j} f(\mu_j).
\end{equation}
The thermodynamic limit of this sum is well known and given by 
\begin{equation}
    \lim_{\rm th} S_0 = \frac{1}{2\pi} \int  d\mu \, f(\mu) , 
\end{equation}
where we have defined the thermodynamic limit $\lim_{\rm th} := \lim_{N,L \to \infty}$. 
After, we have the sum
\begin{equation}
    S_1 = \frac{1}{L} \sum_{\mu_j} \frac{f(\mu_j)}{\mu_j - \lambda_\ell}. 
\end{equation}
We can rewrite it as 
\begin{equation}
    S_1 = \frac{1}{L} \sum_{\mu_j} \frac{f(\mu_j) - f(\lambda_\ell)}{\mu_j - \lambda_\ell} 
    + \frac{1}{2\pi} f(\lambda_\ell) \sum_{j=-\infty}^\infty \frac{1}{j -1/2}.
\end{equation}
Noting that 
\begin{equation}
    \sum_{j=-\infty}^\infty \frac{1}{j -1/2} = 0,
\end{equation}
the thermodynamic limit reads 
\begin{equation}
    S_1 = \frac{1}{2\pi} \mathcal{P} \int d\mu \frac{f(\mu) - f(\lambda_\ell)}{\mu - \lambda_\ell}. 
\end{equation}
Lastly, we evaluate the sum 
\begin{equation}
    S_2 = \frac{1}{L^2} \sum_{\mu_j} \frac{f(\mu_j)}{\left(\mu_j -\lambda_\ell \right)^2}.
\end{equation}
Similarly, it can be written as 
\begin{equation}
    S_2 =\frac{1}{L^2} \sum_{\mu_j} \frac{f(\mu_j)-f(\lambda_\ell)}{\left(\mu_j -\lambda_\ell \right)^2}  +\frac{1}{L^2} \sum_{\mu_j} \frac{f(\lambda_\ell)}{\left(\mu_j -\lambda_\ell \right)^2}. 
\end{equation}
The second term gives 
\begin{equation}
    \frac{1}{L^2} \sum_{\mu_j} \frac{f(\lambda_\ell)}{\left(\mu_j -\lambda_\ell \right)^2} = \frac{1}{4} f(\lambda_\ell),
\end{equation}
where we have used the fact that $\mu_j = 2\pi j/L$ and $\lambda_\ell = 2\pi (\ell+1/2)/L$.
Then, taking the thermodynamic limit of the leading term gives us 
\begin{equation}
    \lim_{\rm th} \frac{1}{L^2} \sum_{\mu_j} \frac{f(\mu_j)-f(\lambda_\ell)}{\left(\mu_j -\lambda_\ell \right)^2} 
    = \frac{1}{2\pi L} \mathcal{P} \int d\mu \frac{f(\mu) -f(\lambda_\ell)}{\mu -\lambda_\ell} 
    = \frac{1}{2\pi L} \del_{\lambda_\ell} \mathcal{P} \int d\mu \frac{f(\mu)}{\mu -\lambda_\ell}.
\end{equation}
Therefore, 
\begin{equation}
    \lim_{\rm th} S_2 = \frac{1}{2\pi L} \del_{\lambda_\ell} \mathcal{P} \int d\mu \frac{f(\mu)}{\mu -\lambda_\ell} + \frac{1}{4} f(\lambda_\ell).
\end{equation}
Consequently, the thermodynamic limit of \eqref{eq.correlator_sum} reduces to 
\begin{equation}\label{eq.correlator_det}
	\mathcal{G}(x,t) = e^{i \eta t} 
	\left(\frac{1}{2\pi} \Upsilon(x,t) + \del_\xi\right) \det(1+\hat{V})\Big|_{\xi=0} 
\end{equation} 
where 
\begin{equation}
	V  = \exp\left\{ -\frac{i}{2}t(q_1 ^2 + q_2 ^2) + \frac{i}{2}x(q_1+q_2)  \right\}
	\sqrt{\rho(q_1)} \left[ \frac{\Xi(q_1)-\Xi(q_2)}{\pi^2 (q_1-q_2)} -\frac{\xi}{2 \pi^3} \Xi(q_1)\Xi(q_2) \right]  \sqrt{\rho(q_2)}
\end{equation}
with
\begin{equation} \label{eq.e_q}
	\Upsilon(x,t) = \int dk \,  e^{itk^2 -ix k} 
	= \sqrt{\frac{\pi}{-it}} \exp\left(-\frac{ix^2}{4t}\right), \quad 	\Xi(x,t,q) = \mathcal{P} \intinf dk \, \frac{e^{it k^2 -ix k}}{k-q} .
\end{equation} 
For practical reasons, we observe that \eqref{eq.correlator_det} is equivalent to 
\begin{equation}\label{eq.Fredholm}
	\avg{\Psi^\dagger(x,t)\Psi(0,0)} = e^{i \eta t} \left[ \det(1+\hat{K}) \left(\frac{1}{2\pi} \Upsilon(x,t) -1\right) +\det(1+\hat{K} -\hat{W}) \right],
\end{equation} 
where 
\begin{equation}
	\begin{aligned}
		K &= e^{ix(q_1+q_2)/2} e^{-it(q_1^2+q_2^2)/2}
        \sqrt{\rho(q_1)} \left[ \frac{\Xi(x,t,q_1)-\Xi(x,t,q_2)}{\pi^2 (q_1-q_2)} \right] \sqrt{\rho(q_2)} \\
		W & = e^{ix(q_1+q_2)/2} e^{-it(q_1^2+q_2^2)/2}
        \sqrt{\rho(q_1)} \left[\frac{1}{2 \pi^3} \Xi(x,t,q_1)\Xi(x,t,q_2) \right] \sqrt{\rho(q_2)} .
	\end{aligned}
\end{equation}

\subsubsection{Regularising $\Xi$}
Consider the function 
\begin{equation} \label{eq.eq_integral}
	\Xi(x,t,q) = \mathcal{P} \int dk \, \frac{e^{itk^2 -ixk}}{k-q} .
\end{equation}
Now, in order to evaluate the integral we observe that \footnote{For simplicity, from now on we omit the symbol $\mathcal{P}$ corresponding to the principal value integral.}
\begin{equation}\label{eq.integral}
\begin{aligned}
    \int\limits_{-\infty}^{+\infty} dk \, \frac{e^{-ixk}e^{itk^2}}{k-q } 
    &= \int\limits_{-\infty}^{+\infty} \frac{dk}{k} \, e^{-ix(k+q)}e^{it(k+q)^2} \\
    &= e^{itq^2-ixq} \int\limits_{-\infty}^{+\infty} \frac{dk}{k} \, e^{-ik(x-2qt)} e^{itk^2} \\
    &:= e^{itq^2-ixq} I(x-2qt,t),
    \end{aligned}
\end{equation}

where 
\begin{equation}
    I (x,t) := \int\limits_{-\infty}^{+\infty} \frac{dk}{k} \, e^{-ikx} e^{itk^2}.
\end{equation}

In order to remove the singularity we differentiate over $x$,    
\begin{equation}
    \del_x I (x,t) = -i \int\limits_{-\infty}^{+\infty} dk \, e^{-ikx} e^{itk^2} = -i \sqrt{\frac{\pi}{-it}} e^{-ix^2/4t} =  \sqrt{\frac{-i \pi}{t}} e^{-ix^2/4t}.
\end{equation}

Integrating back we have 
\begin{equation}
\begin{aligned}
   I (x,t) - I(0,t) = \int_0^x dx\, \del_x I (x,t)  &= \sqrt{\frac{-i \pi}{t}} \int_0^x dx \, e^{-ix^2/4t}  \\
   &= \sqrt{\frac{-i \pi}{t}}\sqrt{\frac{\pi t}{i}} \mathrm{erf}\left(\frac{x\sqrt{i}}{2\sqrt{t}} \right) \\
   & = -i\pi \, \mathrm{erf}\left(\frac{x}{2} \sqrt{\frac{i}{t}} \right) ,
    \end{aligned}
\end{equation}
where 
\begin{equation}
    I (0,t) =  \mathcal{P}\int\limits_{-\infty}^{+\infty} \frac{dk}{k} \, e^{itk^2} = 0.
\end{equation}
Therefore we have that 
\begin{equation}
\begin{aligned}
    I (x,t) &=- i\pi \, \mathrm{erf}\left(\frac{x}{2} \sqrt{\frac{i}{t}} \right).
    \end{aligned}
\end{equation}

Finally, 
\begin{equation}
\begin{aligned}
    \Xi(q,t,x)  &= - i\pi \, e^{itq^2-ixq} \, \mathrm{erf}\left(\frac{x-2qt}{2} \sqrt{\frac{i}{t}} \right).
    \end{aligned}
\end{equation}

\section{Numerical details}\label{app.numerical}

It is worth noting that \eqref{eq.discretisation} is known as the incomplete polylog integral: 
\begin{equation}
    j + \frac{1}{2} = \frac{L}{2\pi} \int_{-\infty}^{\mu_j} \frac{d\lambda}{1+e^{\beta(\lambda^2-\eta)}}  = -\frac{L}{4 \pi \beta} \int_{\sqrt{E_j}}^{\infty} dE \, \frac{E^{-1/2}}{1+e^{E-\beta \eta}} = -\frac{L}{4 \sqrt{\pi} \beta} \mathrm{F}_{-1/2} (\beta \eta,\mu_j),
\end{equation}
where $\mathrm{F}_{-1/2} (\beta \eta,\mu_j) \equiv -\mathrm{Li}_{1/2}(\mu_j,-e^{\beta \eta})$ is the incomplete polylog function. However, we have to rely on numerical methods to solve for the sets of rapidities $\{\bfm\}_{N+1}$ and $\{\bfl\}_N$. Thus, let us consider the points $(x,y) = (j,\mu_j), \, j=0,\dots,N$ which are solutions of eq. \eqref{eq.discretisation}. We have $N+1$ points that provides $N+1$ equations for the $N+1$ coefficients of the polynomial of order $N$  
\begin{equation}
    P_N(x) = \sum_{l=0}^N a_l x^l. 
\end{equation}
The equations are 
\begin{equation}
    \begin{aligned}
        P_N(x_0) &= y_0  \Longrightarrow \sum_{l=0}^N a_l x_0^l  = y_0 \\
        P_N(x_1) &= y_1  \Longrightarrow \sum_{l=0}^N a_l x_1^l  = y_1 \\
        \vdots \\
        P_N(x_N) &= y_N  \Longrightarrow \sum_{l=0}^N a_l x_N^l  = y_N. 
    \end{aligned}
\end{equation}
Thus, the system to be solved is 
\begin{equation}
\begin{pmatrix}
    1 & x_0 & x_0^2 & \dots & x_0^N \\ 
    1 & x_1 & x_1^2 & \dots & x_1^N \\ 
    \vdots & \vdots & \vdots & \ddots &\vdots \\
    1 & x_N & x_N^2 & \dots & x_N^N 
\end{pmatrix}
    \begin{pmatrix}
        a_0 \\
        a_1 \\
        \vdots \\
        a_N
        \end{pmatrix} = 
        \begin{pmatrix}
            y_0 \\
            y_1 \\
            \vdots \\
            y_N
        \end{pmatrix}.
\end{equation}

In terms of our variables, we have that 
\begin{equation}
    \begin{aligned}
        P_N(0) &= \mu_0  \Longrightarrow \sum_{l=0}^N a_l 0^l  = \mu_0 \\
        P_N(1) &= \mu_1  \Longrightarrow \sum_{l=0}^N a_l 1^l  = \mu_1 \\
        \vdots \\
        P_N(N) &= \mu_N  \Longrightarrow \sum_{l=0}^N a_l N^l  = \mu_N, \\
    \end{aligned}
\end{equation}
and the corresponding Vandermonde matrix equation is 
\begin{equation} 
\begin{pmatrix} 
    1 & 0 & 0^2 & \dots & 0^N \\ 
    1 & 1 & 1^2 & \dots & 1^N \\ 
    \vdots & \vdots & \vdots & \ddots &\vdots \\
    1 & N & N^2 & \dots & N^N 
\end{pmatrix} 
    \begin{pmatrix}
        a_0 \\
        a_1 \\
        \vdots \\
        a_N
        \end{pmatrix} = 
        \begin{pmatrix}
            \mu_0 \\
            \mu_1 \\
            \vdots \\
            \mu_N
        \end{pmatrix}.
\end{equation}

\begin{figure}[h!]
    \centering
    \includegraphics{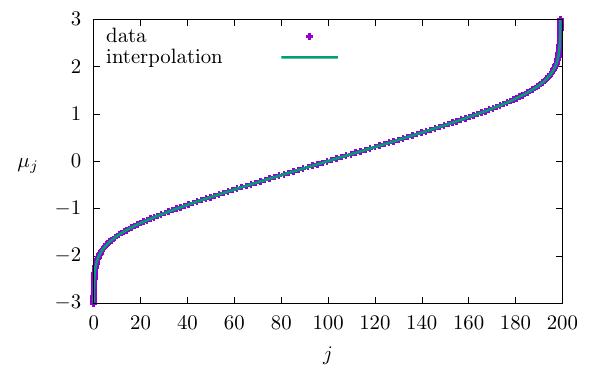}
    \caption{Rapidities distribution in terms of their counting numbers following the interpolation method described.}
    \label{fig.interpol}
\end{figure}

The resulting interpolation procedure is shown in Fig. \ref{fig.interpol}.

Now we turn our attention to the numerical details in order to numerically evaluate Fredholm determinants. The recipe follows the procedures of \cite{Bornemann2009}. The basic ingredients rely on the solution of the linear integral equation
  \begin{equation}
      g(x) + \lambda \int dy \, K(x,y) g(y) = f(x),
  \end{equation}
  which exists only if the infinite rank determinant (the so-called Fredholm determinant) 
  \begin{equation}
      \det(1+\lambda K) = \sum_{n=0}^\infty \frac{\lambda^n}{n!} \int \prod_{j=1}^n dx_j \det(K(x_k,x_\ell))|_{k,\ell=1,\dots,n} \neq 0.
  \end{equation}
We can approximately evaluate it through the introduction of a Gauss-Legendre quadrature of order $\mathcal{O}(N)$ with the set of points and weights $\{x_{j},w_{j}\}|_{j=1,\dots,N}$, namely 
\begin{equation}
    \det(1+K) = \det(\delta_{ij} + \sqrt{w_i} K(x_i,x_j) \sqrt{w_j})|_{i,j=1,\dots,N},
\end{equation}
which is the steps we follow in order to numerically evaluate the correlator in terms of Fredholm determinants described the main text.

	\bibliography{biblio}

\end{document}